\documentclass[aps,noshowkeys]{revtex4}
\usepackage{epsfig}
\usepackage{graphicx}
\usepackage{color}
\usepackage{bm}
\usepackage{amssymb,amsmath,euscript,esint,mathtools}
\begin{document}
\newcommand{\Br}[1]{(\ref{#1})}
\newcommand{\Eq}[1]{Eq.~(\ref{#1})}
\newcommand{\frc}[2]{\raisebox{1pt}{$#1$}/\raisebox{-1pt}{$#2$}}
\newcommand{\frcc}[2]{\raisebox{0.3pt}{$#1$}/\raisebox{-0.3pt}{$#2$}}
\newcommand{\frccc}[2]{\raisebox{1pt}{$#1$}\big/\raisebox{-1pt}{$#2$}}
\newcommand{\RNumb}[1]{\uppercase\expandafter{\romannumeral #1\relax}}
\title{$\mathcal{P}$,~$\mathcal{T}$-odd Faraday rotation in intracavity absorption spectroscopy with molecular beam as a possible way to improve the sensitivity of the search for the time reflection noninvariant effects in nature}
\author{D. V. Chubukov$^{1,2}$, L. V. Skripnikov$^{1,2}$, A.~N.~Petrov$^{1,2}$, V. N. Kutuzov$^1$, L. N. Labzowsky$^{1,2}$}
\affiliation{$^1$ Department of Physics, St. Petersburg State University,
7/9 Universitetskaya Naberezhnaya, St. Petersburg 199034, Russia \\
$^2$Petersburg Nuclear Physics Institute named by B.P. Konstantinov of
National Research Centre ``Kurchatov Institut'', St. Petersburg, Gatchina 188300, Russia 
}

\begin{abstract}
The present constraint on the space parity ($\mathcal{P}$) and time reflection invariance ($\mathcal{T}$) violating electron electric dipole moment ($e$EDM)
is based on the observation of the electron spin precession in an external electric field using
the ThO molecule.
We propose an alternative approach: observation of the $\mathcal{P}$,~$\mathcal{T}$-odd Faraday effect in an external electric field using the cavity-enhanced polarimetric scheme in combination with a molecular beam crossing the cavity. Our theoretical simulation of the proposed experiment with the PbF and ThO molecular beams show that the present constraint on the $e$EDM in principle can be improved by a few orders of magnitude.
\end{abstract}

\maketitle
\section{Introduction}
The existence of the electric dipole moment (EDM) for any particle or closed system of particles violates both the space parity ($\mathcal{P}$) and time-reversal ($\mathcal{T}$) symmetries~\cite{Khrip91,Gin04,Saf18}. Up to date the most stringent experimental constraints for the particles' EDMs are obtained for the electron ($e$EDM) due to its strong enhancement in heavy atoms and diatomic molecules. The most restrictive $e$EDM bounds were established in experiments with the ThO molecule ($|d_e|<1.1\times 10^{-29}$ $e$ cm \cite{ACME18}). Here $e$ is the electron charge. Previously, accurate results were obtained on the Tl atom \cite{Reg02}, YbF molecule~\cite{Hud11} and HfF$^+$ cation~\cite{Cair17}. For extraction of the $e$EDM values from the experimental data, accurate theoretical calculations are required. These calculations were performed for Tl~\cite{Liu92,Dzuba09,Nat11,Por12,Chub18}, for YbF~\cite{Quiney:98, Parpia:98, Mosyagin:98, Abe:14}, for PbF~\cite{Skripnikov:14c,Sudip:15,Chub19:3}, for ThO~\cite{Skripnikov:13c,Skripnikov:15a,Skripnikov:16b,Fleig:16}, and for HfF$^+$~\cite{Petrov:07a,Skripnikov:17c, Fleig:17, Petrov:18b}. In the same experiments it is possible to search for another $\mathcal{P}$,~$\mathcal{T}$-odd effect: $\mathcal{P}$,~$\mathcal{T}$-odd electron-nucleus interaction~\cite{San75,Gor79,Koz95}. Effects originating from this interaction and from $e$EDM can be observed in an external electric field and cannot be distinguished in any particular atomic or molecular experiment. However, they can be distinguished in a series of experiments with different species ( see, e.g. \cite{Bon15,Skripnikov:17c}).

Theoretical predictions of the $d_e$ value are rather uncertain. Within the Standard Model (SM) none of them  promises the $e$EDM value larger than $10^{-38}$ $e$ cm~\cite{Pos14}. However, predictions of the SM extensions are many orders of magnitude larger~\cite{Engel2013}. Different models for the $\mathcal{P}$,~$\mathcal{T}$-odd interactions within the SM framework are discussed in Refs.~\cite{Pos14,Chub16}. In modern experiments for the $\mathcal{P}$,~$\mathcal{T}$-odd effects observation in atomic and molecular systems, either the shift of the magnetic resonance~\cite{Reg02} or the electron spin precession~\cite{Hud11,ACME18,Cair17} in an external electric field is studied.

Due to a large gap between the current experimental bound and the maximum SM theoretical prediction, alternative methods for the observation of the $\mathcal{P}$,~$\mathcal{T}$-odd effects are of interest. In Refs.~\cite{Baran78,Sush78}, it was mentioned the existence of the effect of the optical rotation of linearly polarized light propagating through a medium in an external electric field \--- the $\mathcal{P}$,~$\mathcal{T}$-odd Faraday effect. The possibility of its observation was first studied theoretically and experimentally in Ref.~\cite{Bar88} (see the review on the subject \cite{Bud02}). Recently, a possible observation of the $\mathcal{P}$,~$\mathcal{T}$-odd Faraday effect by the intracavity absorption spectroscopy (ICAS) methods \cite{Boug14,Baev99,Dur10} using atoms was considered \cite{Chub17}. In Ref.~\cite{Boug14} an experiment on the observation of the $\mathcal{P}$-odd optical rotation in the Xe, Hg, and I atoms was discussed. The techniques~\cite{Boug14} are close to what is necessary for the $\mathcal{P}$,~$\mathcal{T}$-odd Faraday effect observation. In Refs.~\cite{Chub18,Chub19:1,Chub19:2} an accurate evaluation of this effect oriented to the application of the techniques~\cite{Boug14} was undertaken for the atomic case and was extended to molecules in Ref.~\cite{Chub19:3}. In the present paper, we consider PbF and ThO for the beam-based ICAS $\mathcal{P}$,~$\mathcal{T}$-odd Faraday effect observation. According to our estimates, these molecules are promising candidate systems for such type of experiment (see below).

As it was shown in earlier works~\cite{Khrip91}, heavy atoms and molecules containing such atoms are promising systems to search for the $\mathcal{P}$,~$\mathcal{T}$-odd effects. For the case of $\mathcal{P}$,~$\mathcal{T}$-odd Faraday effect such systems should also satisfy the following requirements. The natural linewidth of the chosen transitions $\Gamma_{\text{nat}}$ (the collisional width is negligible for beam-based experiments) should be as small as possible, since it allows for the large saturating intensities at large detuning necessary for the $\mathcal{P}$,~$\mathcal{T}$-odd Faraday experiment (see sections~\RNumb{2}-\RNumb{3} below). In other words, it allows reaching a better signal-to-noise ratio in such experiments. For this reason the most suitable are the transition from the ground to the metastable statem, X1$^2\Pi_{1/2}\rightarrow$ X2$^2\Pi_{3/2}$, in the PbF molecule and the transition from the ground to the metastable state X$^1\Sigma_0\rightarrow$ H$^3\Delta_1$ in the ThO molecule. The characteristics of these molecules are discussed in Section~\RNumb{2}. For the molecular case, the applied electric field $\mathcal{E}_{\text{ext}}$ should be close to the saturating field $\mathcal{E}_{\text{sat}}$, which almost completely polarizes a molecule. For diatomic molecules with total electronic angular momentum projection on the molecular axis, $\Omega$, equal to $1/2$, such as PbF, $\mathcal{E}_{\text{sat}}$ is about $10^4$ V/cm. Such a field can be created only within the space of about several centimeters. Diatomic molecules with $\Omega > 1/2$ can be polarized at smaller external electric fields due to closely lying levels of opposite parity (so-called $\Omega$-doubling). The importance of the use of $\Omega$-doubling for the search of $\mathcal{P}$- and $\mathcal{P}$,~$\mathcal{T}$-odd effects was noted in
Refs.~\cite{Lab77,Sush78,Gor79}. 

\begin{figure}[h]
\begin{center}
\includegraphics[width=10.0 cm]{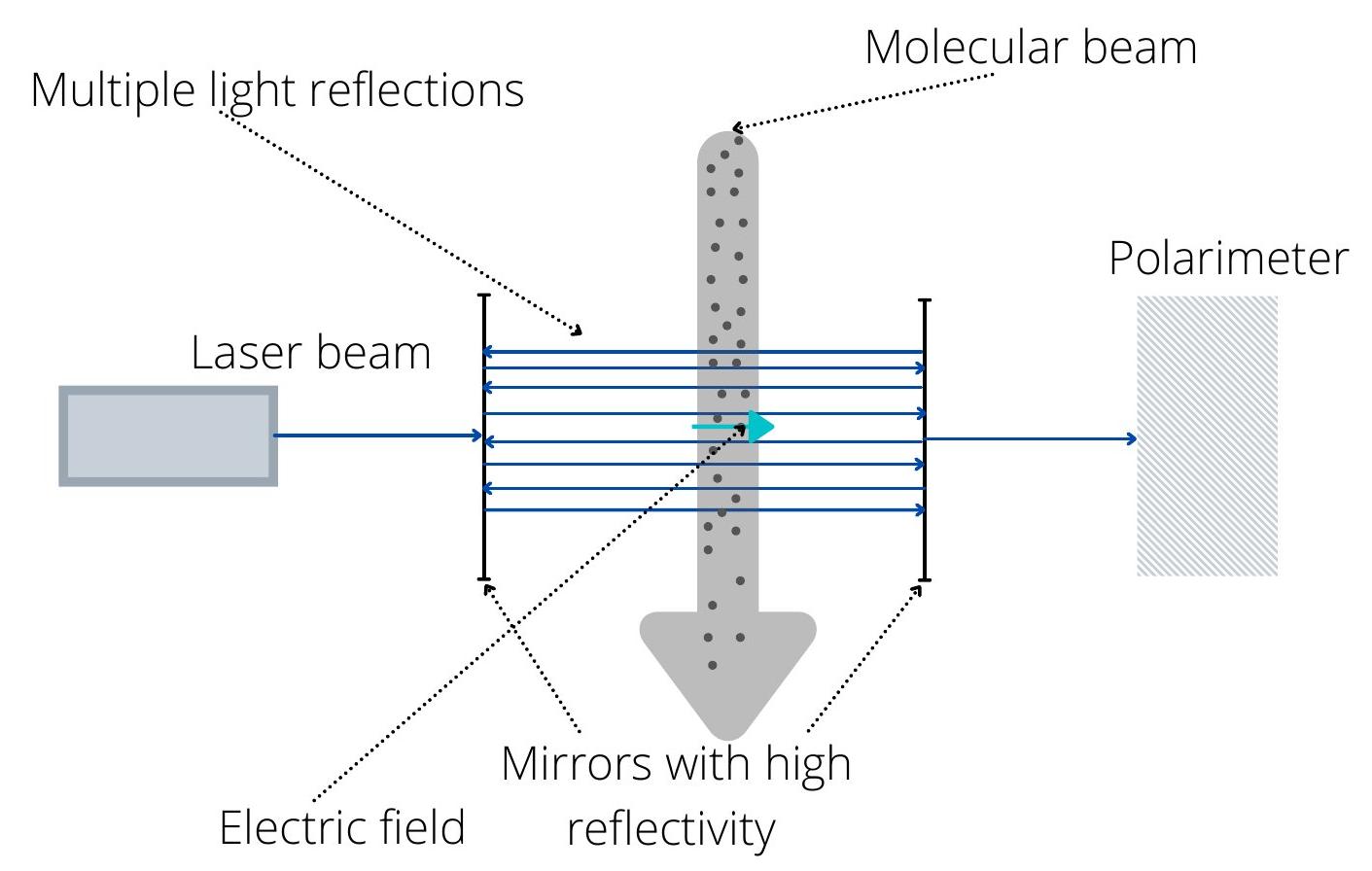}
\end{center}
\caption{\label{f:1} {The principle scheme of the proposed experimental setup. A molecular beam crosses the cavity in a transverse direction. Within the cavity it meets an intracavity laser beam. The crossing of the two beams is located in an electric field oriented along the laser beam. The detection of the optical rotation happens at the output/transmission of the cavity.}}
\end{figure}

We can imagine an ICAS-beam experiment for the $\mathcal{P}$,~$\mathcal{T}$-odd Faraday rotation observation as follows. A molecular beam crosses the cavity in a transverse direction. Within the cavity it meets an intracavity laser beam. The crossing of these two beams is located in an electric field oriented along the laser beam. The detection of optical rotation (either using simple polarimetry, or phase-sensitive techniques) happens at the output/transmission of the cavity (the scheme of the proposed experimental setup is given in Fig.~\ref{f:1}). 

Let us discuss the ultimate ICAS advances necessary for the proposed $\mathcal{P}$,~$\mathcal{T}$-odd Faraday experiments. In Ref.~\cite{Boug14} a possibility to have a total optical path length of about 100~km in a cavity of 1~m length was considered. This results in $10^5$ passes of the light inside the cavity and $10^5$ reflections of the light from the mirrors. For a molecular beam-based experiment with a beam of 1~cm in diameter, typical total optical interaction path-lengths are of about 1~km, i.e. $10^2$ times smaller. However, in another ICAS experiment~\cite{Baev99} an optical path length of $7\times 10^4$~km for a cavity of the same size as in Ref.~\cite{Boug14} was reported. This means that 700 times higher light-pass number inside a cavity may become realistic. Another important property of ICAS experiments is the sensitivity of the rotation-angle measurement. Using a cavity-enhanced scheme a shot-noise-limited birefringence-phase-shift sensitivity at the $3\times 10^{-13}$~rad level was demonstrated~\cite{Dur10}. We consider the above mentioned parameters used in ICAS experiments to assess the realizability of the proposed $\mathcal{P}$,~$\mathcal{T}$-odd Faraday ICAS experiment for the search of the $\mathcal{P}$,~$\mathcal{T}$-odd interactions in molecular physics.

\section{$\mathcal{P}$,~$\mathcal{T}$-odd Faraday experiment on molecules}
The $\mathcal{P}$,~$\mathcal{T}$-odd Faraday effect manifests itself as circular birefringence arising from the light propagating through a medium in an external electric field when the $\mathcal{P}$,~$\mathcal{T}$-odd interactions are taken into account. Its origin is the same as for the ordinary Faraday effect in an external magnetic field. In a magnetic field the Zeeman sublevels split in energy. Then the transitions between two states with emission (absorption) of the right (left) circularly polarized photons correspond to different frequencies since they occur between different Zeeman sublevels. This causes birefringence, i.e. different refractive indices $n^{\pm}$ for the right and left photons. The same happens in an external electric field taking into account the $\mathcal{P}$,~$\mathcal{T}$-odd interactions. In this case, the level splitting is proportional to the linear Stark shift $S^{\Delta}$. The rotation angle $\psi(\omega)$ of the light polarization plane for any type of birefringence looks like
\begin{equation}
 \label{1}
\psi(\omega)=\pi \frac{l}{\lambda} \text{Re} \left[n^+(\omega)-n^-(\omega)\right],
\end{equation}
where $n^{\pm}$ are the refractive indices for the right and left circularly polarized light, $l$ is the optical path length, $\omega$ is the light frequency and $\lambda$ is the corresponding wavelength. In the $\mathcal{P}$,~$\mathcal{T}$-odd Faraday rotation case \cite{Chub18}
\begin{equation}
 \label{2}
\text{Re} \left[n^+(\omega)-n^-(\omega)\right]= \frac{d}{d\omega} \text{Re}\left[n(\omega)\right] S^{\Delta},
\end{equation}
where $n(\omega)$ is the refractive index of linear polarized light.  
In the case of a completely polarised molecule the linear Stark shift of molecular levels is determined by
\begin{equation}
 \label{3}
  S^{\Delta} = d_e \mathcal{E}_{\text{eff}} ,
\end{equation}
where $\mathcal{E}_{\text{eff}}$ is the internal molecular effective electric field acting on the electron EDM. If the molecule is not completely polarized one introduces a corresponding polarization factor that depends on an external electric field $\mathcal{E}_{\text{ext}}$. To extract $d_e$ from the experimental data it is necessary to know the value of $\mathcal{E}_{\text{eff}}$ which cannot be measured and should be calculated (see, e.g., Refs.~\cite{Tit06,Skrip17}).

We evaluated the effective electric fields for the PbF molecule.
The effective electric fields in the PbF molecule were calculated within the relativistic coupled cluster with single, double and noniterative triple cluster amplitudes method using the Dirac-Coulomb Hamiltonian~\cite{Chub19:3}. All electrons were included in the correlation treatment. For Pb the augmented all-electron triple-zeta AAETZ~\cite{Dyall:06} basis set was used. For F the all-electron triple-zeta AETZ~\cite{Dyall:16,Dyall:10,Dyall:12} basis sets were used. The theoretical uncertainty of these calculations can be estimated as 5\%. The value of $\mathcal{E}_{\text{eff}}$ for the ground electronic state is in good agreement with previous studies~\cite{Skripnikov:14c,Sudip:15}. 

The rotation signal $R(\omega)$ in the experiment reads
\begin{equation}
\label{6}
R(\omega)= \frac{\psi(\omega)N_{\text{ev}}}{2 \pi},
\end{equation}
where  $\psi $ is the rotation angle, $N_{\text{ev}}$ is the number of ``events'' in a statistical experiment. In the case under consideration $N_{\text{ev}}$ is the number of photons that had interacted with molecules, and then were detected.
In principle, apart from the losses in the absorber inside the cavity, we have to take into account also the losses in the cavity itself, i.e. in the mirrors. In this work we briefly discuss this part of the losses in section~\RNumb{5}, it changes as a function of intracavity losses and strongly depends on cavity parameters.

Expressed via the spectral characteristics of resonance absorption line the rotation signal reads
\begin{eqnarray}
 \label{8}
R (\omega) & = & \frac{\pi}{3} \frac{l}{\lambda} \rho e^2  |\langle i |  \bm{r}| f \rangle|^2 \frac{h(u,v)}{\hbar\Gamma_D}  \nonumber
\\  & \times &    \frac{2d_e (\mathcal{E}^i_{\text{eff}}+\mathcal{E}^f_{\text{eff}})}{\Gamma_D}N_{\text{ev}},
\end{eqnarray}
\begin{equation}
\label{9}
\omega=\omega_0+ \Delta \omega.
\end{equation}
Here $\rho$ is the molecular number density, $| i \rangle$ and $| f \rangle$
are the initial and final states for the resonance transition, $\bm{r}$ is the electron radius-vector, $\Gamma_D$ is the Doppler width, $\mathcal{E}^i_{\text{eff}}$ and $\mathcal{E}^f_{\text{eff}}$ are the effective fields for the initial and final states, $\omega_0$ is the transition frequency, $\Delta \omega$ is frequency detuning; $\hbar$, $c$ are the reduced Planck constant and the speed of light. \Eq{8} corresponds to the case of E1 resonant transition. For M1 transitions the factor $e^2|\langle i |  \bm{r}| f \rangle|^2$ should be replaced by $\mu_0^2 |\langle i |  \bm{l}-g_S\bm{s}| f\rangle|^2$ where  $\bm{s}$, $\bm{l}$ are the spin and orbital electron angular momenta operators, respectively, $g_S=-2.0023$ is a free-electron $g$ factor and $\mu_0$ is the Bohr magneton. We employ the Voigt parametrization of the spectral line profile~\cite{Khrip91}:
\begin{equation}
\label{11}
 g(u,v) = \text{Im} \; \mathcal{F} (u,v),
\end{equation}
\begin{equation}
\label{12}
f(u,v)= \text{Re}\; \mathcal{F} (u,v) ,
\end{equation}
\begin{equation}
\label{13}
\mathcal{F} (u,v) = \sqrt{\pi} e^{-(u+iv)^2} \left[ 1- \text{Erf} (-i(u+iv)) \right],
\end{equation}
where $\text{Erf}(z)$ is the error function,
\begin{equation}
\label{14}
u=\frac{\Delta\omega}{\Gamma_D},
\end{equation}
and
\begin{equation}
\label{15}
v=\frac{\Gamma_{\text{nat}}}{2\Gamma_D}.
\end{equation}
$\Gamma_{\text{nat}}$ is the natural width. Finally,
\begin{equation}
\label{16}
h(u,v)=\frac{d}{du} g(u,v).
\end{equation}

\begin{figure}[h]
\begin{center}
\begin{minipage}[h]{0.49\linewidth}
\center{\includegraphics[width=6 cm]{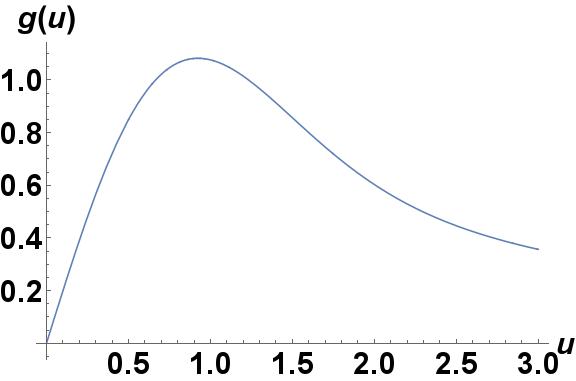}\\ (a)}
\end{minipage}
\hfill 
\begin{minipage}[h]{0.49\linewidth}
\center{\includegraphics[width=6 cm]{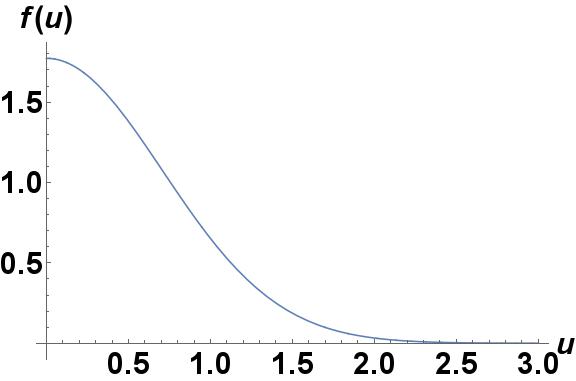} \\ (b)}
\end{minipage}
\vfill
\begin{minipage}[h]{0.49\linewidth}
\center{\includegraphics[width=6 cm]{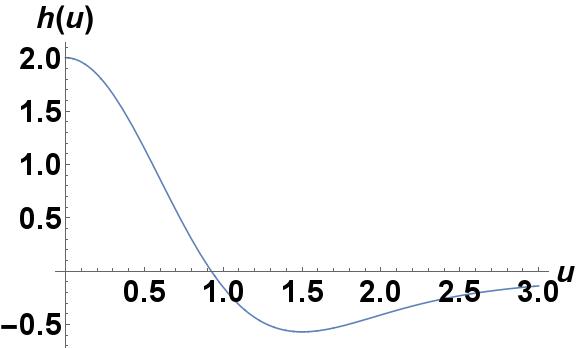} \\ (c)}
\end{minipage}
\renewcommand{\figurename}{Fig.}
\caption{\label{f:4} Behavior of the functions $g(u)$, $f(u)$ and $h(u)$ with $v\ll 1$ close to the resonance: (a) behavior of the rotation angle for optical rotation (natural or $\mathcal{P}$-odd), (b) behavior of the inverse of the absorption length $L$, (c) behavior of the rotation angle for the Faraday effect (ordinary or $\mathcal{P},\mathcal{T}$-odd).}
\end{center}
\end{figure}

The comment can be made on the behavior of the spectral line shape for the considered $\mathcal{P}$, $\mathcal{T}$-odd Faraday effect. The behavior of the functions $g(u)$ and $f(u)$ with $v \ll 1$ is presented in Fig.~\ref{f:4} (a) and Fig.~\ref{f:4} (b), respectively. In Fig.~\ref{f:4} (c), the function $h(u)=\frac{dg}{du}$ with $v \ll 1$ is presented.

 The function $g(u,v)$ defines the behavior of the dispersion, the function $h(u,v)$ determines the behavior of the rotation angle with the detuning. The function $f(u,v)$ defines the behavior of absorption and has its maximum at $\omega_0$. We also introduce $L(\omega)=(\rho \sigma(\omega) )^{-1}$ \--- absorption length with some detuning from the resonance. The cross-section $\sigma(\omega)$ for the photon absorption by a molecule in case of E1 transition looks like 
\begin{equation}
    \label{10}
\sigma(\omega)=4\pi \frac{\omega_0}{\Gamma_D} f(u,v) \frac{e^2|\langle i |\bm{r} | f\rangle |^2}{3\hbar c}.
\end{equation}
 
 Expressed via the absorption length at the arbitrary detuning, the rotation signal reads
 \begin{eqnarray}
 \label{16a}
R (\omega)  =  \frac{h(u,v)}{f(u,v)} \frac{l}{L(u,v)}  \frac{d_e (\mathcal{E}^i_{\text{eff}}+\mathcal{E}^f_{\text{eff}})}{2\Gamma_D}N_{\text{ev}}.
\end{eqnarray}
 
  The maximum of $h(u,v)$ also coincides with $\omega_0$. However, it has a second maximum~\cite{Chub18}, which allows to observe the $\mathcal{P}$,~$\mathcal{T}$-odd Faraday effect off-resonance, in the region where absorption is small. In the following we choose $\Delta\omega=5\Gamma_{D}$. At this detuning the absorption drops down essentially ($f(u,v)\sim v/u^2$), but the rotation is still close to its second maximum ($h(u,v)\sim 1/u^2$). 
  Here we do not consider the hyperfine structure. If the hyperfine structure is resolved, it does not change the order-of-magnitude estimate for the rotation angle. However, the choice of certain hyperfine levels depends on particular experiment. 

\begin{table} [h!]
\caption{Parameters of transitions under investigation in molecular species. The adopted number density for different species is $\rho\sim 10^{10}$~cm$^{-3}$.}
\tabcolsep=0.01cm
\scalebox{0.9}{\begin{tabular}{cccccc}
\hline\hline  Molecule & Transition & Wavelength  & Linewidth  & Effective field & Absorption length    \\
&  & $\lambda$, nm & $\Gamma_{\text{nat}}$, s$^{-1}$ & $\mathcal{E}_{\text{eff}}$, GV/cm & $L(u=5)$, cm \\
\hline 
PbF & X1 $^2\Pi_{1/2} \rightarrow$ X2 $^2\Pi_{3/2}$ & 1210 & $2.7 \times 10^3$  & 
38.0(X1), 9.3(X2) &
 $2\times 10^{9}$  \\
ThO & X $^1\Sigma_0\rightarrow$  H $^3\Delta_1$ & 1810 & $5\times 10^{2}$ & 
0(X), 80(H) &
  $1\times 10^{10}$  \\
\hline \hline
\end{tabular}}
\label{table:1}
\end{table}

1) One of promising candidates for the ICAS $\mathcal{P}$,~$\mathcal{T}$-odd Faraday experiment with diatomic molecules is the PbF molecule with the X1 $^2\Pi_{1/2} \rightarrow$ X2 $^2\Pi_{3/2}$ transition ($\lambda=1210$~nm). The natural linewidth of the X2 state is $\Gamma_{\text{nat}}=2.7 \times 10^3$~s$^{-1}$~\cite{Das02}. For PbF beam we adopt the transverse temperature of 1~K (e.g., in Ref.~\cite{Alm17} the transverse temperature of the supersonic YbF beam was reported to be about 1~K) and the transverse $\Gamma_D= 4.5 \times 10^{7}$~s$^{-1}$. Our calculations give the following effective electric fields values: $\mathcal{E}_{\text{eff}} (^2\Pi_{+1/2}) = 38$~GV/cm and $\mathcal{E}_{\text{eff}} (^2\Pi_{+3/2}) = 9.3$~GV/cm. One can adopt the achievable number density of PbF molecules approximately as $\rho\sim 10^{10}$~cm$^{-3}$. Then, according to \Eq{10}, the absorption length at dimensionless detuning $u=5$, $L(u=5)\sim 2\times 10^{9}$~cm. In Table~\ref{table:1} the parameters of the transition under investigation in PbF are listed.

2) Consider the X $^1\Sigma_0\rightarrow$ H $^3\Delta_1$ transition ($\lambda=1810$~nm) in ThO. This transition lies in the infrared region. It is interesting to consider such a molecular system since the best constraint on the $e$EDM was obtained on ThO. The natural linewidth of the metastable H state is $\Gamma_{\text{nat}}=5\times 10^{2}$~s$^{-1}$~\cite{ACME18}. The effective electric field for the H state was calculated in~\cite{Skripnikov:13c,Skripnikov:15a,Skripnikov:16b,Fleig:16}. For the ThO beam ($T=1$~K) the transverse $\Gamma_D= 2.9 \times 10^{7}$~s$^{-1}$. In Ref.~\cite{ACME14} the number density of ThO molecular beam was reported to be about $\rho\sim (10^{10}-10^{11})$~cm$^{-3}$. We adopt the number density of ThO molecules as $\rho\sim 10^{10}$~cm$^{-3}$. Then, according to \Eq{10}, the absorption length at dimensionless detuning $u=5$, $L(u=5)\sim 1\times 10^{10}$~cm. In Table~\ref{table:1} the parameters of the transition under investigation in ThO are listed.

In the following sections we will investigate theoretically in more detail the ICAS-beam $\mathcal{P}$,~$\mathcal{T}$-odd Faraday experiment on the PbF and ThO molecules with the intensities near the saturation threshold.

\section{Shot-noise limit and saturation limit}

The signal (R) to noise (F) ratio in this section we will write via the number of ``events'' $N_{\text{ev}}$:
\begin{equation}
\label{17}
\frac{R}{F}=\frac{\psi N_{\text{ev}}}{2\pi \sqrt{N_{\text{ev}}}}=  \frac{\psi\sqrt{N_{\text{ev}}}}{2\pi} .
\end{equation}
Here $\psi$ is the rotation angle of light polarization plane.
The number $N_{\text{ev}}$ in the $\mathcal{P}$,~$\mathcal{T}$-odd Faraday experiment should be defined as a number of photons which have interacted with molecules and then detected. The total number of photons $N_{\text{phot}}$ involved in the experiment may be larger than the number of involved molecules $N_{\text{mol}}$, may be smaller than $N_{\text{mol}}$, may be equal to it. We will be interested in the case when $N_{\text{phot}}\gg N_{\text{mol}}$.

For shot-noise limited measurement, the condition $\frac{R}{F} > 1$ should be fulfilled. One way of collecting statistics is a continuous experiment with a cw laser. The other way is to collect statistics in many sets (e.g. with a pulsed laser) if the condition $\frac{R}{F} > 1$ is not fulfilled during one set of the experiment. Nevertheless, after repeating the experiment $n$ times the statistically improved signal-to-noise ratio
 \begin{equation}
\label{17d}
\frac{R}{F}=\frac{\psi\sqrt{nN_{\text{ev}}}}{2\pi},
\end{equation}
in principle, can be made arbitrary large.
This means that the shot-noise limited measurement without observation of the angle $\psi$ in any particular measurement, in principle, is also possible. In this case one should collect the statistics from many measurements. The same way of collecting statistics is used in the ACME experiments~\cite{ACME18}.

For the shot-noise limited measurement we have to make the number $N_{\text{ev}}$ (i.e. the number of photons) as large as possible. However, this number is limited by the saturation effects. In section~\RNumb{4} we show that for suggested experiments we do not reach this limit.

For laser beams of high intensity the laws of nonlinear optics should be applied. The refractive index $n(\omega)$ depends on the intensity of the light $I(\omega)$ in the following way~\cite{Boyd}:
\begin{equation}
\label{18}
n(\omega)=\frac{n_0(\omega)}{1+I(\omega)/I_{\text{sat}}(\omega)},
\end{equation}
where $n_0(\omega)$ is the refractive index for weak light and $I_{\text{sat}}(\omega)$ is the saturation intensity. When the light intensity exceeds the saturation one, $I(\omega)> I_{\text{sat}}(\omega)$, both absorption and dispersion decrease. Equation~\Br{18} is derived within the two-level model of an atom (a molecule) which is valid for the resonant processes of our interest.

It is instructive to look at \Eq{18} from the point of view of Einstein relations between the spontaneous and stimulated emission~\cite{Ber83}:
\begin{equation}
\label{20}
W_{if}^{\text{st}}=W_{fi} = \frac{\pi^2c^2}{\hbar \omega^3} J(\omega) W_{if}^{\text{sp}}, 
\end{equation}
where  $W_{if}^{\text{sp}}$ is the spontaneous probability (transition rate) for transition between the initial ($i$) and final ($f$) states (which can be approximated as natural linewidth for the transition $\Gamma_{\text{nat}}$), $W_{if}^{\text{st}}$ stands for stimulated emission and $W_{fi}$ corresponds to the absorption probability. Equation~\Br{20} is written for the polarized anisotropic (laser beam) radiation with frequency $\omega$, $J(\omega)d\omega=I(\omega)$. The dimensionless coefficient at $W_{if}^{\text{sp}}$ defines the ``number of photons in the field'' $N$. When a certain transition  
$i\rightarrow f$ is considered, $d\omega\sim \Gamma_{\text{nat}}$. Then the number $N$ defines actually the relative importance of the spontaneous and stimulated emission. If $N<1$, the spontaneous emission dominates, for $N>1$ the stimulated emission dominates. The condition $I=I_{\text{sat}}$ in \Eq{18} according to Ref.~\cite{Boyd} corresponds to $N\approx 1$ in \Eq{20}, i.e. the saturation intensity can be obtained from the condition $W_{if}^{\text{st}}\approx W_{if}^{\text{sp}}$.

Taking into account the detuning from the resonance and the Doppler width, one can also represent the stimulated emission and absorption probabilities in terms of the absorption cross-section as it was done, for instance, in Ref.~\cite{Sieg86}:
\begin{equation}
\label{21}
W_{if}^{\text{st}}=W_{fi} = \frac{\sigma(\omega)I(\omega)}{\hbar \omega}.
\end{equation}

Then, the saturation intensity which reduces the refractive index down to one-half can be expressed as follows:
\begin{equation}
\label{22}
I_{\text{sat}}= \frac{\hbar \omega}{\sigma \tau_s},
\end{equation}
where $ \tau_s$ is the saturation time constant (or the effective lifetime or the recovery time). It is the time for the molecules to become excited and to decay again. This time can be approximated as $ \tau_s\approx \left(\Gamma_{\text{nat}}\right)^{-1}$. As it was noted in Ref.~\cite{Sieg86}, from Eqs.~\Br{21}-\Br{22} clear physical meaning of the saturation intensity follows. It means one photon incident on each atom or molecule, within its cross-section $\sigma$, per the recovery time $ \tau_s$.

Substituting the absorption cross-section from \Eq{10} to \Eq{22} one obtains the expression for the saturation intensity:
\begin{equation}
\label{23}
I_{\text{sat}}(\omega,u)= \frac{\hbar \omega^3 \Gamma_D}{\pi c^2 f(u,v)}.
\end{equation}

The most important feature is that for any intensity $I\geqslant  I_s$ the effect of saturation does not arise instantaneously and takes the saturation time $t_{\text{sat}} \sim \left(W_{if}^{\text{st}} \right)^{-1}$ for its formation. For the off-resonance measurement $t_{\text{sat}}$ can be large enough.

It is interesting to compare the resonance and large-detuned cases in terms of signal-to-noise ratio. Then, the figure-of-merit is as follows. One should consider the next ratio:
\begin{equation}
\label{26b}
\frac{\psi(u=5) \sqrt{I_{\text{sat}}(u=5)}}{\psi(u=0) \sqrt{I_{\text{sat}}(u=0)}}\sim  \frac{1}{v}  \frac{l}{L(u=5)} \sqrt{\frac{u^2}{v}}.
\end{equation}
For the case of ThO,  $v\sim 10^{-5}$, $L(u=5)\sim 10^5$~km  (for $\rho\sim 10^{10}$~cm$^{-3}$). Then, for two existing cavities with achievable effective optical pathlnegths:

1) $l=1$~km, according to \Eq{26b}, the ratio~$\sim 10^3$;

2) $l=700$~km, according to \Eq{26b}, the ratio~$\sim 10^6$.

It follows that large detuned case has a great advantage over the resonance one.

\section{ICAS-beam experiment with the number of photons larger than the number of molecules}

For large detunings in an ICAS-beam experiment one can have the number of photons much larger than the number of molecules (as long as $I/I_{\text{sat}}\lesssim 1$), so a medium (molecules) is by no means continuous and the saturation effects are of importance. The Beer-Lambert law and the standard optimization ($l=2L(\omega)$~\cite{Khrip91}) are no longer valid. To zeroth order, we can set  $N_{\text{ev}} \approx N_{0}$ in \Eq{6} where $N_{0}$ is the initial number of photons injected into the cavity. According to \Eq{6} and \Eq{8}, the expression for the $\mathcal{P}$,~$\mathcal{T}$-odd Faraday rotation angle can be presented as: 
\begin{equation}
    \label{24}
    \psi(\omega)= (\rho l) \left[ \text{cm}^{-2} \right] K \left[ \frac{\text{cm}}{e} \right] d_e \left[e\;\text{cm} \right].
\end{equation}
For the X1 $^1\Pi_{1/2}\rightarrow$ X2 $^3\Pi_{1/2}$ transition in the PbF molecule, $ K \approx 2 \times 10^3 \;\text{cm}/e $. 
In the scenario employed in this paper there is no optimal condition $l=2L(\omega)$. In principle, the optical pathlength is limited only by the quality of the mirrors in a cavity. For $\rho\sim 10^{10}$~cm$^{-3}$, $d_e \approx 1.1 \times 10^{-29}$~$e$~cm and optical pathlength $l=1$~km (corresponding to cavity~\cite{Boug14} intersected by a molecular beam of 1~cm in diameter), according to \Eq{24}, one obtains $ \psi\sim 2\times 10^{-11}$~rad. Under the same conditions but with optical pathlength $l=700$~km (corresponding to cavity~\cite{Baev99} intersected by a molecular beam of 1~cm in diameter), according to \Eq{24}, one obtains $ \psi\sim 10^{-8}$~rad. Another thing one should worry about is that the experiment cannot last more than the saturation time. However, in our scenario not the time of experiment but the transit time of a molecule through the laser beam plays a key role. 

Let us estimate the saturation intensity for the transition under investigation in the PbF molecule. According to \Eq{23}, for $\omega=1.56\times 10^{15}$ s$^{-1}$, $\Gamma_D=4.5\times10^7$ s$^{-1}$, $\Gamma_{\text{nat}}=2.7\times 10^3$ s$^{-1}$ and $u=5$ one gets
\begin{equation}
\label{25}
I_{\text{sat}}(u=5)=5.3 \times 10^3 \;\; \text{W/cm}^2.
\end{equation}
Such an intensity corresponds to the injection of $N\sim 3\times 10^{20}$ photons per second through a laser beam cross-section of 1 mm$^2$. Taking into account \Eq{20}, such a saturation intensity corresponds to the case when $W_{if}^{\text{st}}=W_{fi} \approx \Gamma_{\text{nat}}=2.7\times 10^3 \;\; \text{s}^{-1}$. 

The next question is: how many PbF molecules inside the crossing volume are in the excited (X2 $^2\Pi_{3/2}$) state if the laser intensity is equal to the saturation one? For simplicity and without loss of generality we consider the following statement of the problem and do not consider any technical issues. The PbF molecular beam of 1~cm in diameter travels through a cavity of 1~m length in a transverse direction with the speed $v_{\text{mol}} \approx 300$~m/s. Continuous laser light of 1~mm in diameter of the saturation intensity is coupled to the cavity. Then the transit time of the PbF molecule to pass through the laser beam is $\tau_{\text{tr}}\approx 10^{-5} $~s. The fraction of the molecules in the excited state for the case when $W_{fi}\tau_{\text{tr}} \ll 1$ can be estimated as
\begin{equation}
\label{26}
(1-e^{-W_{fi}\tau_{\text{tr}}})\approx W_{fi}\tau_{\text{tr}}=\Gamma_{\text{nat}}\tau_{\text{tr}}\approx 0.03.
\end{equation}
It means that if the saturation intensity is coupled to the cavity then only 3\% of the total number of PbF molecules in the crossing volume will be in the excited state.  

Alternatively, one can define the saturation parameter $\kappa=$excitation rate$(u)/$relaxation rate.   The excitation rate$(u)$ is proportional to the intensity $I$. At the detuning the excitation rate $(u=5)/$  excitation rate $(u=0)$   
scales as$\sim f(u,v)/ f(0,v)\sim v/u^2$ (at the considered conditions $v/u^2$ is a small number). The choice of the saturation intensity $I_{\text{sat}}$ corresponds to $\kappa=1$. In this case one has $\sim 33\%$ of molecules in the excited state and $\sim 67\%$ of molecules in the ground state. It means that one does not ``bleach'' the molecules. However, in our proposal, since $1/\tau_{\text{tr}}>$relaxation rate, we should define the saturation parameter as $\kappa=$excitation rate$(u)\cdot\tau_{\text{tr}}$. As a result, in such a beam-based ICAS experiment, the number of detected photons can be increased by several orders of magnitude. 

For the ICAS-beam experiment with the ThO molecules, the coefficient $K$ in \Eq{24} is $K\approx 4\times 10^3\;\text{cm}/e$. Substituting the adopted parameters of ThO ($\omega=1.04\times 10^{15}$~s$^{-1}$, $\Gamma_D=2.9\times10^7$~s$^{-1}$, $\Gamma_{\text{nat}}=5\times 10^2$~s$^{-1}$ and $u=5$) in \Eq{23}, one obtains  
\begin{equation}
\label{26a}
I_{\text{sat}}(u=5)=3.5 \times 10^3 \;\; \text{W/cm}^2.
\end{equation}
This value corresponds to the injection of $N\sim 3\times 10^{20}$ photons per second through the laser cross-section of 1~mm$^2$. Estimating, similarly to the the PbF case, the fraction of the molecules in excited state $\Gamma_{\text{nat}} \tau_{\text{tr}}\approx 0.005$. That is, if the saturation intensity is coupled to the cavity then only 0.5\% of the total number of ThO molecules in the crossing volume will be in the excited state. This makes it possible to increase the intensity coupled to the cavity by an order of magnitude. In this case, one has $N\sim 3\times 10^{21}$ photons per second through the laser cross-section $\sim$ 1 mm$^2$ and 5\% of the total number of ThO molecules in the excited state in the crossing volume.

The next question concerns fundamental noises which determine the statistical error of the experiment. The figure-of-merit for the fundamental noise-limited experiment on the molecular spin-precession observation (ACME-style) is as follows:
\begin{equation}
\label{26c}
\delta d_e \sim  \frac{1}{\mathcal{E}_{\text{eff}}}\frac{1}{\tau_{\text{coh}}} \frac{1}{\sqrt{\dot{N}_{\text{mol}}T}},
\end{equation}
where $\tau_{\text{coh}}$ is the coherence time (a few ms), $\dot{N}_{\text{mol}}$ is the number of molecules supplied to the experiment by the molecular beam per unit time in the desired initial molecular state and $T$ is the time of the experiment. In the ACME experiment the statistics is determined by the number of molecules (molecular spin-noise). The experiment is carried out on the excited state of a molecule with nonzero total angular momentum (spin).

Contrary to this, in our proposed experiment we do not need to prepare molecules in the excited state. Our experiment is carried out with molecules in the ground (zero spin) states.  In the ground state of the ThO molecule, we don't have any spin. In turn, excited by a laser state is optically inactive to this laser (provided decoherence effects are negligible). So there is no molecular spin noise in this case. Note, that taking into account the nuclear spin, the PbF molecule can also formally be considered as a molecule with zero total angular momentum in the ground state \cite{Rav08}. The excited states are produced in small amounts ($\sim$1\%)  during the experiment and also they have no spin noise. So statistics in our case is defined by the number of detected photons. This number can be much larger than that of molecules due to large detuning. 
The figure-of-merit for the fundamental noise-limited experiment in our proposed large detuned case is as follows:
\begin{equation}
\label{26d}
\delta d_e \sim  \frac{1}{\mathcal{E}_{\text{eff}}} \frac{1}{\tau_{\text{coh}}} \frac{\Gamma_{\text{nat}} L(u)}{c} \frac{1}{\sqrt{\dot{N}_{\text{phot}}T}},
\end{equation}
where $\dot{N}_{\text{phot}}$ is the number of detected photons per unit time and $\tau_{\text{coh}}=l/c$ is the coherence time in the optical rotation experiment. The factor $N_{\text{phot}}$ is the key difference between our proposal and the ACME-style experiments.

\section{Cavity transmission in the ICAS-beam experiment}

In this section we consider the cavity transmission, $T_{\text{cav}}$, i. e. the transmission of the light determined by the properties of the mirrors. In principle, the sensitivity of the ICAS experiments strongly depends on the parameters of mirrors. In this paper we only briefly consider the problem of the cavity transmission in the simplest model.

Let us consider two identical mirrors with high reflectivity $R=1-\delta$, $\delta\approx (10^{-5}-10^{-7})$. Transmission of such an interferometer can be described as \cite{Mes}:
\begin{equation}
\label{27}
T_{\text{cav}}=\frac{I_{\text{tr}}}{I_{\text{in}}}\approx \frac{4(1-R)^2(1-A/2)}{(2-2R+A)^2},
\end{equation}
where $I_{\text{tr}}$ is the transmitted intensity, $I_{\text{in}}$ is the initial laser intensity and $A$ is the light intensity loss in the absorber during one round trip. For the case of PbF beam for one pass ($l\sim 1$~cm) and a detuning $u=5$, $A
<\rho \sigma  l\sim 10^{-9}$. Then, $A\ll \delta$ and, in principle, can be negligible. In this case, $I_{\text{tr}} \approx I_{\text{in}}$. Note also, that it is possible to choose such an initial laser intensity, that the coupled intracavity intensity $I_{\text{int}}=I_{\text{tr}}/\delta$ will reach the saturation intensity. This means that the transmitted intensity is now $I_{\text{tr}}=I_{\text{sat}}\times \delta$.
The photon shot-noise limit for an ideal polarimeter (see, e.g., the review~\cite{Bud02}) is
\begin{equation}
\label{28}
\delta\psi\approx \frac{1}{2\sqrt{N_{\text{phot}}}},
\end{equation}
where $N_{\text{phot}}$ is the number of detected photons.

Consider two cases of the existing cavities:

1a) The cavity~\cite{Boug14} with $\delta\sim 10^{-5}$, according to \Eq{25}, for the PbF case gives $I_{\text{tr}}=I_{\text{sat}}\times \delta\approx 5.3\times 10^{-2}$~W/cm$^2$. It corresponds to the detection of $N_{\text{phot}}\sim 3\times 10^{15}$ photons per second. Then, in such experiments with the integration time of the order of two weeks $\sim 10^6$~s (such an observation time was in the ACME experiments), the number of detected photons is $N_{\text{phot}}\sim 3\times 10^{21}$. According to \Eq{28}, in this case $\delta\psi\sim 10^{-11}$~rad. According to \Eq{24} for such a cavity and the recent ACME experimental bound on the $e$EDM value, $\psi \sim 10^{-11}$~rad. As a result, for such parameters PbF is a candidate to verify the recent ACME results via the alternative method.

1b) The cavity~\cite{Boug14} with $\delta\sim 10^{-5}$, for the ThO case, gives $N\times \delta\approx 3\times 10^{21}\times 10^{-5}\approx 3\times 10^{16}$ detected photons per second. Then, in such experiments with the integration time on the order of two weeks $\sim 10^6$~s, the number of detected photons is $N_{\text{phot}}\sim 3\times 10^{22}$. According to \Eq{28}, in this case $\delta\psi\sim 3\times 10^{-12}$~rad. According to \Eq{24} for such a cavity and the recent ACME experimental bound on the $e$EDM value, $\psi \sim 4 \sim 10^{-11}$~rad. As a result, ThO is a good candidate for improving the $e$EDM bound by 1~order of magnitude. 

2a)  The cavity~\cite{Baev99} with $\delta\sim 10^{-7}$, according to \Eq{25},  for the PbF case gives $I_{\text{tr}}=I_{\text{sat}}\times \delta\approx 5.3\times 10^{-4}$~W/cm$^2$. It corresponds to the detection of $N_{\text{phot}}\sim 3\times 10^{13}$ photons per second. Then, in such experiments with the integration time on the order of two weeks $\sim 10^6$~s, the number of detected photons is $N_{\text{phot}}\sim 3\times 10^{19}$. According to \Eq{28}, in this case $\delta\psi\sim 10^{-10}$~rad. According to \Eq{24} for such a cavity and the recent ACME experimental bound on the $e$EDM value, $\psi \sim 10^{-8}$~rad. As a result, PbF is a good candidate for improving the $e$EDM bound by 2~orders of magnitude in this case. 

2b)  The cavity~\cite{Baev99} with $\delta\sim 10^{-7}$, for the ThO case, gives $N\times \delta\approx 3\times 10^{21}\times 10^{-7}\approx 3\times 10^{14}$ detected photons per second. Then, in such experiments with the integration time on the order of two weeks $\sim 10^6$~s, the number of detected photons is $N_{\text{phot}}\sim 3\times 10^{20}$. According to \Eq{28}, in this case $\delta\psi\sim 3\times 10^{-11}$~rad. According to \Eq{24} for such a cavity and the recent ACME experimental bound on the $e$EDM value, $\psi \sim 3 \times 10^{-8}$~rad. As a result, ThO is a good candidate for improving the $e$EDM bound by 3~orders of magnitude in this case.

In conclusion of the section, we comment on the possible sources of improving the $\mathcal{P}$,~$\mathcal{T}$-odd Faraday signal-to-noise ratio. Note, that according to \Eq{8}, the rotation angle is proportional to $\psi \sim h(u,v)/\Gamma_D^2$. For large dimensionless detunings $u$ (e.g., $u=5$), $\psi\sim 1/(u \Gamma_D)^2$. For the case where the number of photons is much larger than the number of molecules (near the saturation threshold), one can neglect the absorption of photons. Then, it is no longer necessary to make such a large detuning. However, even at the detuning $u=1.5$ (the second extremum of the $h(u,v)$ function) where $h(u,v) \approx 0.5$,  $\psi\sim (1/5) \times 1/\Gamma_D^2$. Thus, the rotation angle enhances by a factor of $~5$, but the shot-noise (connected in our case with the saturation intensity \Eq{23} which depends on $u$) drops down by a factor of $\sim \sqrt{5}$. As a result, such smaller detuning can improve the signal-to-noise ratio by no more than a factor of two. Obviously, increasing the number density of molecular beam and increasing of the optical pathlength (the quality of the mirrors) lead to improving the signal-to-noise ratio. Note also, the value of the $\mathcal{P}$,~$\mathcal{T}$-odd Faraday rotation is determined, among other things, by the largest of the widths (natural, collisional, transit-time, Doppler, etc.). For instance, for the PbF molecular beam case, the largest width is the Doppler one ($\Gamma_{\text{nat}}=2.7 \times 10^3$~s$^{-1}$, the transit-time width $\Gamma_{\text{tr}}\sim 1/(2\pi\tau_{\text{tr}})\approx 1.6\times 10^4$~s$^{-1}$, $\Gamma_D= 4.5 \times 10^{7}$~s$^{-1}$).  Finally, we would like to mention that with squeezed states of light the photon shot-noise limit can be surpassed which would be favorable for the $\mathcal{P}$,~$\mathcal{T}$-odd Faraday effect observation. However, these squeezed states of light have not yet found their application in polarimetry.

\section{Conclusions}

The recent most advanced $e$EDM constraint obtained in the experiment with ThO is $|d_e|<1.1\times 10^{-29}$ $e$ cm. In this experiment, electron spin precession in an external electric field is employed and the effect is proportional to the time spent by a particular molecule in an electric field. In the present paper we suggest another method for observation of such effects \--- a beam-ICAS $\mathcal{P}$,~$\mathcal{T}$-odd Faraday experiment with molecules. A theoretical simulation of the proposed experiment is based on the recently available ICAS parameters. In this experiment it is not necessary to keep a separate molecule in an electric field since the effect is accumulated in the laser beam which encounters many molecules. According to our estimates for the PbF molecule, the current $e$EDM sensitivity can be improved by 1-2 orders of magnitude. In its turn, for the ThO molecule the current $e$EDM sensitivity can be improved by 1-3 orders of magnitude. This implies testing of new particles at energy 1-2 order of magnitude larger than the current best constraint.

Making these predictions we understand that some technical problems, not mentioned here, may arise. In this paper we did not discuss the possible systematic errors among which the stray magnetic fields, the uncontrolled ellipticity of the laser beam and uncontrolled drift of mirrors are the most evident. A problem of avoiding the $\mathcal{P}$-odd optical rotation, much stronger than the $\mathcal{P}$,~$\mathcal{T}$-odd rotation also should be resolved. All these problems we hope to address in the future studies.

\begin{acknowledgments}
Preparing the paper and the calculations of the $\mathcal{P}$,~$\mathcal{T}$-odd Faraday signals, as well as finding optimal parameters for experiment were supported by the Russian Science Foundation grant 17-12-01035. L.V.S. acknowledges the support of the Foundation for the advancement of theoretical physics and mathematics ``BASIS'' grant according to the research projects No.~18-1-3-55-1. D.V.C. acknowledges the support of the President of Russian Federation Stipend No.~SP-1213.2021.2.
The authors would like to thank Dr. Dmitry Budker, Dr. Mikhail G. Kozlov, Dr. L. Bougas, Dr. Timur A. Isaev and  Dr. Peter Rakitzis for helpful discussions.
\end{acknowledgments}


\begin{thebibliography}{99}
\section*{References}
\expandafter\ifx\csname natexlab\endcsname\relax\def\natexlab#1{#1}\fi
\expandafter\ifx\csname bibnamefont\endcsname\relax
  \def\bibnamefont#1{#1}\fi
\expandafter\ifx\csname bibfnamefont\endcsname\relax
  \def\bibfnamefont#1{#1}\fi
\expandafter\ifx\csname citenamefont\endcsname\relax
  \def\citenamefont#1{#1}\fi
\expandafter\ifx\csname url\endcsname\relax
  \def\url#1{\texttt{#1}}\fi
\expandafter\ifx\csname urlprefix\endcsname\relax\def\urlprefix{URL }\fi
\providecommand{\bibinfo}[2]{#2}
\providecommand{\eprint}[2][]{\url{#2}}





\bibitem{Khrip91}
I.B. Khriplovich, Parity Nonconservation in Atomic Phenomena, Gordon and Breach, London, 1991

\bibitem{Gin04}
J.S. Ginges and V.V. Flambaum, Phys. Rep. \textbf{397}, 63 (2004)


\bibitem{Saf18}
M.S. Safronova, D. Budker, D. DeMille, D.F.J. Kimball, A. Derevianko, C.W. Clark, Rev. Mod. Phys. \textbf{90}, 025008 (2018)


\bibitem{ACME18}
V. Andreev et al. (ACME collaboration), Nature \textbf{562}, 355 (2018)


\bibitem{Reg02}
B.C. Regan, E.D. Commins, C.J. Schmidt and D. DeMille, Phys. Rev. Lett. \textbf{88}, 071805 (2002)

\bibitem{Hud11}
J.J. Hudson, D.M. Kara, I.J. Smallman, B.E. Sauer,
M. R. Tarbutt and E. A. Hinds, Nature \textbf{473}, 493 (2011)


\bibitem{Cair17}
W.B. Cairncross, D.N. Gresh, M. Grau, K.C. Cossel, T.S. Roussy, Y. Ni, Y. Zhou, J. Ye and E.A. Cornell, Phys.Rev.Lett. \textbf{119}, 153001 (2017)


\bibitem{Liu92}
Z.W. Liu and Hugh P. Kelly, Phys. Rev. A \textbf{45}, R4210(R) (1992)

\bibitem{Dzuba09}
V.A. Dzuba and V.V. Flambaum, Phys.Rev. A \textbf{80}, 062509 (2009)

\bibitem{Nat11}
H.S. Nataraj, B.K. Sahoo, B.P. Das, and D. Mukherjee, Phys. Rev. Lett. \textbf{106}, 200403 (2011)

\bibitem{Por12}
S.G. Porsev, M.S. Safronova and M.G. Kozlov, Phys.Rev.Lett. \textbf{108}, 173001 (2012)

\bibitem{Chub18}
D.V. Chubukov, L.V. Skripnikov and L.N. Labzowsky, Phys. Rev. A \textbf{97}, 062512 (2018)

\bibitem{Quiney:98}
H.M. Quiney, H. Skaane, and I.P. Grant, J.Phys. B: At.Mol.Opt.Phys.  \textbf{31}, 85 (1998)
 
\bibitem{Parpia:98}
F. Parpia, J.Phys. B: At.Mol.Opt.Phys. \textbf{31}, 1409 (1998)

\bibitem{Mosyagin:98}
N.S. Mosyagin, M.G. Kozlov, and A.V. Titov, J.Phys. B: At.Mol.Opt.Phys. \textbf{31}, L763 (1998)


\bibitem{Abe:14}
M. Abe, G. Gopakumar, M. Hada, B.P. Das, H. Tatewaki, and D. Mukherjee, Phys. Rev. A \textbf{90}, 022501 – (2014)

\bibitem{Skripnikov:14c}
L.\,V. Skripnikov, A.\,D. Kudashov, A.\,N. Petrov, A.\,V. Titov, Phys.Rev. A \textbf{90}, 064501 (2014)


\bibitem{Sudip:15}
S. Sasmal, H. Pathak, M.\,K. Nayak, N. Vaval, S. Pal J. Chem. Phys. \textbf{143}, 084119 (2015)

\bibitem{Chub19:3}
D.~V.~Chubukov, L.~V.~Skripnikov, L.~N.~Labzowsky, JETP Letters \textbf{110}, 382 (2019) [Pis'ma v ZhETF \textbf{110}, 363 (2019)]


\bibitem{Skripnikov:13c}
L.V. Skripnikov, A.N. Petrov, and A.V. Titov, J.Chem.Phys. \textbf{139}, 221103 (2013)
   
\bibitem{Skripnikov:15a}
L.V. Skripnikov, and A.V. Titov, J.Chem.Phys. \textbf{142}, 024301 (2015)
  
 \bibitem{Skripnikov:16b}
L.V. Skripnikov, J.Chem.Phys. \textbf{145}, 214301 (2016)

\bibitem{Fleig:16}
M. Denis, and T. Fleig, J.Chem.Phys. \textbf{145}, 214307 (2016)

\bibitem{Petrov:07a}
A.N. Petrov, N.S. Mosyagin, T.A. Isaev, and A.V. Titov, Phys.Rev. A \textbf{76}, 030501(R) (2007)

\bibitem{Skripnikov:17c}
L.V. Skripnikov, J.Chem.Phys. \textbf{147}, 021101 (2017)
 
\bibitem{Fleig:17}
 T. Fleig, Phys.Rev. A \textbf{96}, 040502 (2017)
 
\bibitem{Petrov:18b}
A.N. Petrov, L.V. Skripnikov, A.V. Titov, and V.V. Flambaum, Phys.Rev. A \textbf{98}, 042502 (2018)



\bibitem{San75}
P.G.H. Sandars, At.Phys. \textbf{4}, 71 (1975)

\bibitem{Gor79}
V.G. Gorshkov, L.N. Labzowsky and A.N. Moskalev, ZhETF \textbf{76}, 414 (1979)
[Sov. Phys. JETP \textbf{49}, 209 (1979)]

\bibitem{Koz95}
M.G. Kozlov, and L.N. Labzowsky, J.Phys. B: At.Mol.Opt.Phys. \textbf{28}, 1933 (1995)

\bibitem{Bon15}
A.A. Bondarevskaya, D.V. Chubukov, O.Yu. Andreev, E.A. Mistonova, L.N. Labzowsky, G. Plunien, D. Liesen, F. Bosch, J. Phys. B \textbf{48}, 144007 (2015)

\bibitem{Pos14}
M. Pospelov and A. Ritz, Phys. Rev. D \textbf{89}, 056006 (2014)

\bibitem{Engel2013}
J. Engel, M. J. Ramsey-Musolf, and U. van Kolck, Prog. Part. Nucl. Phys. \textbf{71}, 21 (2013)



\bibitem{Chub16}
D.V. Chubukov and L.N. Labzowsky, Phys. Rev. A \textbf{93}, 062503 (2016)

\bibitem{Baran78}
N.B. Baranova, Yu.V. Bogdanov and B.Ya. Zel'dovich, Usp.Fiz.Nauk \textbf{123}, 349 (1977) [Sov.Phys.Usp. \textbf{20}, 1977, (1978)]

\bibitem{Sush78}
O.P. Sushkov,and V.V. Flambaum, ZhETF \textbf{75}, 1208 (1978) [Sov. Phys. JETP \textbf{48}, 608 (1978)]

\bibitem{Bar88}
L.M. Barkov, M.S. Zolotorev and D.A. Melik-Pashaev, JETP letters \textbf{48}, 144 (1988)

\bibitem{Bud02}
D. Budker, W. Gawlik, D. Kimball, S.M. Rochester, V. Yashchuk and A. Weis, Rev.Mod.Phys. \textbf{74}, 1153 (2002)

\bibitem{Boug14}
L. Bougas, G.E. Katsoprinakis, W. von Klitzing and T.P. Rakitzis, Phys. Rev. A \textbf{89}, 052127 (2014)

\bibitem{Baev99}
V.M. Baev, T. Latz and P.E. Toschek, Appl. Phys. B \textbf{69}, 171 (1999) 

\bibitem{Dur10}
M. Durand, J. Morville and D. Romanini, Phys. Rev. A \textbf{82}, 031803 (2010)


\bibitem{Chub17}
D.V. Chubukov and L.N. Labzowsky, Phys. Rev. A \textbf{96}, 052105 (2017)

\bibitem{Chub19:1}
D.V. Chubukov, L.V. Skripnikov, L.N. Labzowsky, V.N. Kutuzov, and S.D. Chekhovskoi, Phys.Rev. A \textbf{99}, 052515 (2019)

\bibitem{Chub19:2}
D.V. Chubukov, L.V. Skripnikov, V.N. Kutuzov, S.D. Chekhovskoi, and L.N. Labzowsky, Atoms, 7, 56; doi:10.3390/atoms7020056 (2019)



\bibitem{Lab77}
L.N. Labzovsky, Zh. Eksp. Teor. Fiz. \textbf{73}, 1623 (1977)
[Sov. Phys. JETP \textbf{46}, 853 (1977)]



\bibitem{Tit06}
A.V. Titov, N.S. Mosyagin, A.N. Petrov, T.A. Isaev and D. DeMille, ``Study of parity violation effects in polar heavy-atom molecule'', In: Recent Advances in the Theory of Chemical and Physical Systems ["Progress in Theoretical Chemistry and Physics" series] v.15, chapt.2, pp. 253-284 (2006)


\bibitem{Skrip17}
L.V. Skripnikov, J.Chem.Phys., \textbf{147}, 021101 (2017)


\bibitem{Dyall:06}
K.G. Dyall, Theor. Chem. Acc.,  \textbf{115}, 441 (2006)


\bibitem{Dyall:16}
K.G. Dyall, Theor. Chem. Acc.,  \textbf{135}, 128 (2016)


\bibitem{Dyall:10}
A.S.P. Gomes, K.G. Dyall and L. Visscher, Theor. Chem. Acc. \textbf{127}, 369 (2010)

\bibitem{Dyall:12}
K.G. Dyall, Theor. Chem. Acc. \textbf{131}, 1217 (2012)




\bibitem{Das02}
K.K. Das, I.D. Petsalakis, H.-P. Liebermann, A.B. Alekseyev, R.J. Buenker, J.Chem.Phys. \textbf{116}, 608 (2002)



\bibitem{Alm17}
J.R. Almond, PhD Thesis ``Laser cooling of YbF molecules for an improved measurement of the electron electric dipole moment'', Imperial College London (2017)


\bibitem{ACME14}
J. Baron et al., Science \textbf{343}, 269–272 (2014)

\bibitem{Boyd}
R. Boyd, ``Nonlinear Optics'',  Academic Press (2008)


\bibitem{Ber83}
V.B. Berestetskii, E.M. Lifshits and L.P. Pitaevskii, Quantum Electrodynamics (Oxford: Pergamon) 1982  

\bibitem{Sieg86}
A.E. Siegman, ``Lasers'', University Science Books Mill Valley, California (1986)

 \bibitem{Rav08}
 C.P. McRaven, P. Sivakumar, and N.E. Shafer-Ray, Phys.Rev. A \textbf{78}, 054502 (2008).

\bibitem{Mes}
D. Meschede, Optics, Light and Lasers, Wiley-VCH 2007




\end{thebibliography}
\end{document}